\relax
\documentstyle [art12]{article}
\begin {document}
\bibliographystyle {plain}

\title{\bf One-Dimensional Kondo Lattice at Half Filling: Charge Properties.}
\author {A. M. Tsvelik}
\maketitle
\begin {verse}
$Department~ of~ Physics,~ University~ of~ Oxford,~ 1~ Keble~ Road,$
\\
$Oxford,~OX1~ 3NP,~ UK$\\
\end{verse}
\begin{abstract}
\par
The model of Kondo chain with $M$-fold degenerate
band of conduction electrons of spin 1/2 at half filling
interacting with localized
spins $S$ is studied. It is shown that due to the instanton effects
the spin sector reveals itself in optical conductivity. Namely,
the density of the
charge current at large
wave vector $k \approx \pi$ is proportional to the density of
topological charge of the staggered magnetization. As a consequence, the
optical conductivity at $k = \pi$ would either have a gap equal
to three spin gaps (for  $|M - 2S| = $(even) when the spin spectrum is
gapful), or show a psedogap $\sigma(\omega, k = \pi) \sim \omega^2$
for $|M - 2S|$ = (odd).
\end{abstract}

\sloppy
\par
 In this letter I continue to study  the model of one dimensional
of Kondo chain at half filling introduced in the previous publication
$^1$. This model  has the following Hamiltonian:

\begin{eqnarray}
H = \sum_r \sum_{\alpha = 1}^M
[ - \frac{1}{2}(c^+_{r + 1,\alpha, a} c_{r,\alpha, a}
+ c^+_{r,\alpha, a} c_{r + 1,\alpha, a})
+ J (c^+_{r,\alpha, a}\hat{\vec\sigma}_{ab}c_{r,\alpha, b})
\vec S_r] \label{eq:model}
\end{eqnarray}
The conduction electrons have spin 1/2 and their band has an
additional $M$-fold degeneracy. The value of local spins is S.

 It was shown in Ref. 1 that
the low energy behavior of the Kondo chain with a  half filled
electron band is determined by antiferromagnetic correlations
(the RKKY coupling) and the Kondo effect plays no role.
Depending on the value of $2S - M$ the spin excitations are
either gapless (if $2S - M$ is odd) or have a spectral gap
exponentially small in $t/J$ ($2S - M$ is even). Thus  the
spin correlation length is always large and the low energy sector
can be described by a continuous field  theory.
It was shown that this theory is the $O(3)$ nonlinear sigma model
with the topological term. The coefficient of the topological term
is $\pi$(mod $2\pi$)  for odd  $2S - M$ and 0 (mod $2\pi$) for
even $2S - M$. These results are in qualitative agreement with
the strong coupling picture and with the numerical results of
other authors$^{2,3}$.

 In the previous publication I made some remarks about the charge
sector of the model. In particular I argued that the charge gap was
equal to the spin gap and provided the explicit expression for the
wave function of an electron-kink bound state. These arguments were
justly critisized by X.-G. Wen who pointed out that the suggested
wave function is not normalizable$^4$.

 In this letter I present rigorous results for  the charge sector of
of the model (~\ref{eq:model}). It turns out that despite of the
fact that the charge excitations do have a large gap $\Delta_{ch}
\approx JS$, the spin sector contributes into
the optical conductivity at $k = \pi$ through instantons. This
contribution is purely dynamical. For even $2S - M$
the indirect gap measured by optical conductivity
is equal to $3\Delta_{sp}$ ($\Delta_{sp}$ is the spin gap) and
for odd $2S - M$ there is a only a pseudogap.

 In order to analyse charge properties of the model (1), we have to
recall some arguments of Ref. 1. The suggested approach is based on
the fact that at $J/t << 1$ the spin subsystem is very slow.
This allows to treat  the staggered magnetization $\vec n$ as a
slow variable and
linearize the fermionic spectrum around the Fermi points
\begin{eqnarray}
\vec m_r = a\vec k(x) + (-1)^r\vec n(x)\sqrt{1 - a^2\vec k(x)^2},
(\vec k \vec n) = 0 , \nonumber\\
c_r = i^r\psi_R(x) + (-i)^r\psi_L(x) \label{eq:separ}
\end{eqnarray}
Substituting this expression into the initial model (1) we get the
following Lagrangian density for fermions:
\begin{eqnarray}
L =  \bar\psi_j
[i\gamma_{\mu}\partial_{\mu}\hat I + JS(\hat{\vec\sigma}\vec n(x))
\sqrt{1 - a^2\vec k(x)^2}]\psi_j \label{eq:action1}
\end{eqnarray}
where the term coupled to $\vec k$ is omitted as irrelevant. The
corresponding fermionic determinant was calculated in Ref. 1.  In what
follows we shall be interested in the expansion of this determinant
in gradients of $\vec n$. The coordinate dependence of the amplitude
of the mass term coming from the $\sqrt{1 - a^2\vec k(x)^2}$-term
is not important in this respect. Therefore we shall treat this term
as constant: $JS<\sqrt{1 - a^2\vec k(x)^2}> = m$. It
was shown that this determinant was a particular case of the more
general one:
\begin{eqnarray}
D[g] = MTr\ln[i\gamma_{\mu}\partial_{\mu} +
(1 + i\gamma_S)mg/2 + (1 - i\gamma_S)mg^+/2] = MW(g) \label{eq:ident}\\
W(g) =  \int d^2x
\{\frac{1}{16\pi}Tr(\partial_{\mu}g^+\partial_{\mu}g)
+ \frac{1}{24\pi}
\int_0^1 d\xi \epsilon_{abc}Tr(g^+\partial_a gg^+\partial_b gg^+
\partial_c g)\}
\label{eq:Det}
\end{eqnarray}
where $g$
belongs to the SU(2) group, $\gamma_S = i\sigma_3$  and $W(g)$ is the famous
Wess-Zumino-Witten action. The equivalence becomes obvious
when one makes  the transformation
\begin{equation}
\psi_R \rightarrow i\psi_R, \psi_L \rightarrow \psi_L
\end{equation}
and put $n_0 = 0$ in the general expression for the matrix $g$:
\begin{equation}
g = n_0 + i(\vec\sigma\vec n)\sqrt{1 - n_0^2}
\end{equation}

 The derivation of Eq.(~\ref{eq:ident}) given in Ref. 1. can be
simplified. Namely, one can use the procedure of non-Abelian
bosonization (see Refs. 5 - 7, I also recommend the book by Itzykson
and Drouffe$^8$). In this procedure the action of free
fermions on a group U(2) (this can be done for any group,
but I do not want to complicate the arguments) can be rewritten as
follows:
\begin{eqnarray}
\int d^2x i\bar{\psi}_a\gamma_{\mu}\partial_{\mu}\psi_a = W(h) +
\frac{1}{4\pi}\int d^2x(\partial_{\mu}\phi)^2,\\
\psi^+_{R,a}\psi_{L,b} \sim e^{i\phi}h_{ab}
\end{eqnarray}
where $h$ belongs to the SU(2) group and $W$ is given by
Eq.(~ref{eq:Det}). Using this equivalence we get the following:
\begin{eqnarray}
S = \int d^2x [i\bar{\psi}_a\gamma_{\mu}\partial_{\mu}\psi_a +
\frac{m}{2}\bar{\psi}_a(1 + i\gamma_S)g_{ab}{\psi}_a + \nonumber\\
\frac{m}{2}\bar{\psi}_a(1 - i\gamma_S)g^+_{ab}\psi_b] = \nonumber\\
\frac{1}{4\pi}\int d^2x(\partial_{\mu}\phi)^2 +  W(h) + \tilde m\int
d^2x Tr(
e^{i\phi}g^+h + H.c)
\end{eqnarray}
Now we do the transformation $g^+h = G$ and use the well-known
identity
\begin{equation}
W(gG) = W(G^+) + W(g) + \frac{1}{16\pi}\int d^2x(g^+\partial_{\bar z}g
G\partial_z G^+)
\end{equation}
As a result we get
\begin{eqnarray}
S = S_1 + W(g)\\
S_1 = \frac{1}{4\pi}\int d^2x(\partial_{\mu}\phi)^2 + W(G^+) + \tilde
m\int d^2x Tr(
e^{i\phi}G + H.c) \label{eq:der}
\end{eqnarray}
Since the operator $e^{i\phi}G$ has the right and left
conformal dimensions $(1/2,
1/2)$, it is relevant and the theory described by the action $S_1$ is
massive with the mass gap $\tilde m \approx JS$.
The remaining part of the action is equal to $W(g)$ which reproduces
the result Eq.(~\ref{eq:ident}).

 Now we shall use the identity (~\ref{eq:ident}) to calculate the
correlation function of electronic currents. For this purpose we
calculate the variational derivative of  Eq.(~\ref{eq:ident}) with
respect to $n_0$ at $n_0 = 0$ to establish the following Ward identity:
\begin{eqnarray}
J_S = m\bar\psi\gamma_S\psi = \frac{\delta D(g)}{\delta n_0}|_{n_0 = 0} =
\nonumber\\
- \frac{M}{4\pi}\frac{\delta }{\delta n_0}\int d^2x Tr[\delta g
g^+\partial_{\mu}(g^+\partial_{\mu}g)] =
\frac{iM}{\pi}\epsilon_{\mu\nu}\left(\vec n[\partial_{\mu}\vec
n\times\partial_{\nu}\vec n]\right) \label{eq:topcur}
\end{eqnarray}
Thus the anomalous fermionic current is proportional to the density of
topological charge of the staggered magnetization. In this way the
charge and the spin sector are related to each other.

 In terms of the original fermions the anomalous current is nothing
but the ordinary current at $k = \pi$. Therefore the pair correlation
function of $J_S$ is proportional to the optical conductivity at $k =
\pi$:
\begin{equation}
\sigma(\omega,k = \pi - q) = \frac{e^2}{\omega}\Im m<<J_S(-\omega, -
q)J_S(\omega, q)>>
\end{equation}

 In the case $2S - M$ = (even) the spin sector is described by the
ordinary O(3) nonlinear sigma model.  The spin sector has a gap and  the
correlation function of topological densities decay exponentially in
Euclidean  space-time. It was shown by Kirillov and Smirnov$^{9}$
that the first non-vanishing
formfactor in the Lehmann expansion of this correlation function
contains three elementary excitations, that is
the gap measured by the optical conductivity
is equal to $3\Delta_{sp}$. The case $2S - M$ = (odd)
is more interesting. In this case the spin sector is critical and
is described by the O(3) nonlinear sigma model with $\theta = \pi$.
The latter is equivalent to the S = 1/2 Heisenberg chain, that is can
be written as a theory of the massless scalar bosonic field $\Phi$:
\begin{equation}
S_{\pi} = \frac{1}{2}\int d^2x(\partial_{\mu}\Phi)^2
\end{equation}
The components of the staggered magnetization
are expressed as exponents of $\Phi$ and the dual
field $\Theta$ defined as
\begin{equation}
\epsilon_{\mu\nu}\partial_{\nu}\Theta = \partial_{\mu}\Phi \label{eq:theta}
\end{equation}
namely$^{10}$,
\begin{equation}
\vec n \sim \left(\cos(\sqrt{2\pi}\Phi),
\cos(\sqrt{2\pi}\Theta),\sin(\sqrt{2\pi}\Theta)\right)
\end{equation}

 Substituting this expression into Eq.(~\ref{eq:topcur}) and using the
identity (~\ref{eq:theta}) we get
\begin{equation}
J_S \sim \sin(\sqrt{2\pi}\Phi)(\partial_{\mu}\Phi)^2
\end{equation}
The pair correlation function of the currents is
\begin{equation}
<<J_S(-\omega_n, -q)J_S(\omega_n, q)>> \sim (\omega_n^2 +
v^2q^2)^{3/2}
\end{equation}
and the optical conductivity
\begin{equation}
\sigma(\omega, k = \pi - q) \sim \frac{1}{\omega}\Im m[\omega^2 -
v^2q^2]^{3/2}\end{equation}
where $v$ is the velocity of spin waves. The latter means that the
optical conductivity would reveal in this case only a pseudogap at $k
= \pi$.

  As we see, from one side it is fair to say that the spectrum of the
system is separated into two sectors (we call them the charge and the
spin sectors) and the charge sector always has a large gap
$\Delta_{ch}$
of order of $JS$ in agreement with the strong coupling expansion and
the results of Refs. 2, 3. This conclusion follows from the discussion around
Eq.(~\ref{eq:der}). It does not mean however, that optical
measurements measure only $\Delta_{ch}$ (they do it only at small wave
vectors).  Due to the fact that
the physical fermions are composite fields, there is the admixture
of spin sector. The very form of the identity (~\ref{eq:topcur})
tells us that this admixture originates from instantons, which is not
unusual$^{10}$.

  I am
grateful to X.-G. Wen  and M. Sigrist for the constructive criticism,
for I. Lerner and to A. Ludwig for the interest to the work.


\end{document}